\documentclass[aps,prb,twocolumn,showpacs,amsmath,amssymb]{revtex4}
\usepackage{epsfig}

\begin{document}
\title{Proximity effect gaps in S/N/FI structures}
\author{Daniel Huertas-Hernando$^{1,2}$, Yu. V. Nazarov$^{2}$}
\affiliation{$\emph{1}$ Center for Theoretical Physics, Sloane Physics Laboratory, 
Yale University,New Haven, CT 06520, USA.\\ 
$\emph{2}$ 
Kavli Institute of NanoScience Delft, 
Delft University of Technology, 
Lorentzweg 1, 2628 CJ,Delft, The Netherlands
}
\date{\today}

\begin{abstract} 
We study the proximity effect in
hybrid structures consisting of superconductor
and ferromagnetic insulator separated by 
a normal diffusive metal (S/N/FI structures).
These stuctures were proposed to realize the
absolute spin-valve effect. We pay special attention
to the gaps in the density of states of the normal
part. We show that the effect of the ferromagnet 
is twofold: It not only shifts the density of states
but also provides suppression of the gap. 
The mechanism of this suppression is remarkably similar
to that due to magnetic impurities.
Our results are obtained from the solution of
one-dimensional Usadel equation
supplemented with boundary conditions for matrix current
at both interfaces. 
\end{abstract}


\maketitle

\section{Introduction}

The research on heterostrustures that combine superconductors
and ferromagnets has begun in sixties.\cite{DeGennes}
Still, the $F/S$ structures
remain a subject of active experimental and theoretical investigation.
New developments concern Josephson, $\pi$-junctions,\cite{pi} 
spin valves based on giant magnetoresistance effect,\cite{spinvalve} 
triplet superconducting ordering, \cite{triplet} Andreev reflection phenomena in $S-F$ 
trilayers\cite{trilayers} and proposed detection of electron entanglement.\cite{entanglement}

Near the $F/S$ interface electrons are influenced by 
both exchange field $h$ of the ferromagnet and pair
potential $\Delta$ of the superconductor.  
Exchange field tends to split the density of states
so that the energy bands for different spin directions are shifted
in energy.\cite{RefTedrowPRL} Besides, the exchange field at $F/S$ interfaces 
induces pair breaking, suppression of the spectrum gap and even 
formation of a superconducting gapless state.\cite{Gapless}
The latter is qualitatively similar to the gapless state
in superconductors with  paramagnetic impurities.\cite{AG}
The exchange field is also active if the ferromagnet 
is an insulator (FI): Although in this case electrons
can not penetrate the ferromagnet,
they pick up the exchange field while reflecting from
the $FI/S$ interface. \cite{WeertArnold,RefTokuyasu}
This has been experimentally verified with 
EuO-Al$\vert$ Al$_{2}$O$_{3}$ $\vert$ Al junctions.
\cite{RefKumarPRL}

The physics of $F/S$ or $FI/S$ structures is thus governed
by two factors:
i. electron states are modified by $\Delta$ and $h$,
ii. the $\Delta$ and $h$ are modified as a result of
the modification of electron states by virtue of self-consistency equations. 
$\Delta$ and $h$ correspond to incompatible types of ordering
that suppress each other and therefore compete rather than collaborate. 
It was suggested \cite{RefDaniPRLGBC} 
that $S/N/FI$ structures can be used  
to get rid of the second factor.
The buffer normal metal effectively 
separates $\Delta$ and $h$ in space to prevent their mutual
suppression, provided its thickness exceeds the superconducting
coherence length. However, the electrons in the normal do feel both 
superconducting and ferromagnetic correlations.
Varying the conductances of the $S/N$ and $N/FI$ interfaces  
it is possible to tune the strength of these correlations.

If there are no ferromagnetic correlations, the traditional
proximity effect in $S/N$ structures takes place.
A $S/N$ interface couples electrons and holes in the normal metal 
by the coherent process of Andreev reflection
\cite{AndreevINTRO} at energies $\varepsilon \simeq \Delta $,
\cite{TinkhamINTRO} so that Andreev bound states are formed.\cite{KulikINTRO} If the normal metal
is connected to the bulk superconductor only, 
there is a (mini) gap in the spectrum
of these states. The energy scale of the gap
is not $\Delta$. Rather, it is set by the inverse
escape time into the superconductor, $\hbar/\tau_{E} \ll \Delta$.
This minigap was first 
predicted in Ref. \onlinecite{RefMcMillanPRL} and has been intensively
investigated.\cite{RefMcMillan1PRL} A common realistic assumption
is that diffusive transport takes place in the normal metal.
\cite{PaulSDF}
In this case, the superconducting
proximity effect 
is described by the Usadel equation. \cite{UsadelFDS}

The $S/N/FI$ structures  can
be used to achieve an \emph{absolute spin-valve effect}.
\cite{RefDaniPRLGBC}
The collaboration of superconducting and 
ferromagnetic correlations results in a spin-split BCS-like 
DOS in the normal metal part,
very much like in a BCS superconductor
in the presence of the spin magnetic field. \cite{RefTedrowPRL}
However, the effective exchange field  $\tilde{h}$ and proximity gap $\tilde \Delta$
characterizing the DOS in this case,\cite{RefDaniPRLGBC} are parametrically different
from $h$ and $\Delta$ in the ferromagnet and superconductor. 
In particular, the fact that the 
effective exchange field $\tilde{h}$ affects electrons in all points of the normal 
part of a $S/N/FI$ structure, 
does not imply to that the ``real'' exchange field $h$ in the ferromagnet 
penetrates into $N$
 by an appreciable distance. Actually it is known that $h$ only penetrates up 
to distances of the Fermi wavelength in the normal part. Rather, the effect 
of $\tilde{h}$ is 
due to the extended nature of the electron wave functions in $N$, which probe 
the spin dependent potential at the $FI/N$ interface and carry the information 
about this potential throughout the 
whole normal metal region.  

Two such stuctures with normal metal parts connected by
a tunnel junction constitute the absolute spin valve.\cite{RefDaniPRLGBC}
The presence of a normal metal is essential to provide a
physical separation between the sources of superconducting and magnetic
correlations so that superconductivity and magnetism do not
compete. The fact that the magnet is an insulator guarantees the
absence of the normal electrons at the Fermi level and thus enables the
proximity gap. The spin-valve effect mentioned would not be absolute if the ferromagnet is a metal. In this case, there would be a finite density of states at any energy due to the possible electron escape into the ferromagnet. Besides, the use of a insulator does not require  nearly fully spin polarized ferromagnets (half-metal materials) to achieve an absolute spin-valve effect.

The analysis of Ref. \onlinecite{RefDaniPRLGBC} was
restricted to the so-called ``circuit-theory" limit.\cite{RefYuliFDS}
The variation of Green functions along the normal part 
was disregarded. This is justified if the 
resistance of the normal metal part is much smaller than both the
resistance of the $S/N$ interface and effective spin-mixing
resistance characterizing the $N/FI$ interface.

In the present paper, we extend this analysis to
arbitrary resistances of the diffusive normal metal part.
To do so, we analyse the solutions of one-dimensional 
Usadel equation\cite{UsadelFDS} in the normal part. 
Our goal is to find the gaps in the spectrum for both spin directions.
The equation must be supplemented by boundary conditions at
both magnetic and superconducting interfaces.

Microscopic models for interfaces of mesoscopic structures 
have been extensively studied in past years combining the 
scattering matrix approach and quasiclassical 
Green's function theory.\cite{RefZaitsev, RefSauls, 
RefYuliFDS, RefBCKopu} In the case of diffusive conductors 
the interfaces can be described in a compact/transparent 
way by means of "circuit theory" boundary 
conditions.\cite{RefYuliFDS} In that case the interface 
is described by a set of conductance parameters given as 
certain specific combinations of the reflection and/or 
transmission amplitudes of the scattering matrix associated 
with the interface. The regime of diffusive transport considered is distinct
from the (quasi)ballistic regime assumed in many studies of
$F/S$ structures. \cite{WeertArnold, RefTokuyasu, Zareyan}

To summarize the resuls shortly, we have shown that the effect
of magnetic correlations is twofold. Firstly, 
these correlations
shift the BCS-like densities of states in energy, with shifts 
being opposite for opposite spin directions.\cite{RefTedrowPRL,
RefTokuyasu,RefDaniPRLGBC} 
In the first approximation,
the absolute value of the proximity 
gap is not affected by the ferromagnetic insulator.
Secondly, we have also found that 
the magnetic correlations may suppres the gap.
The gap completely dissapears at some critical values of the parameters.
This is qualitatively different from Ref. \onlinecite{RefDaniPRLGBC} and 
presents the effect of the resistance of the normal metal.

The mechanism of the gap suppression appears to be surprisingly similar
to that due to paramagnetic impurities. \cite{AG} 
At qualitative level, this has been noticed in the context of 
$FI/S$ structures. \cite{WeertArnold,RefTokuyasu} 
However, for $S/N/FI$ structures the analogy becomes closer: the gap
closing in a certain limit (see Section \ref{sec:4SDF}) 
is described by equations identical to those of Ref. \onlinecite{AG}. 
We stress that there are no magnetic impurities in our model
and the quasiparticles are affected by magnetism only
when they are reflected at the $N/FI$ boundary.
Effective spin-flip time thus arises from interplay of magnetic
correlations and diffusive scattering in the normal metal. 
As we have already mentioned, the spectrum gap suppression
in $S/N/FI$ structures is not accompanied by suppression
of the pair potential, this is in distinction from the situation 
described in.\cite{AG,WeertArnold,
RefTokuyasu}

The structure of the article is as follows. In section \ref{sec:1SDF}
we introduce the basic equations
and define the matrix current in the diffusive
normal metal.
 In section \ref{sec:2SDF}, we specify the boundary conditions for the
Usadel equation by imposing matrix current conservation at the interfaces
of our 
$S/N/FI$ system. 
The resulting set of equations allows us to calculate the total
Green's function $\check{G}$ at any point in the diffusive wire. 
Analytical solutions can be only found in two
limiting cases (sections \ref{sec:3SDF} and \ref{sec:4SDF}). 
Further on, we
solve numerically the 
equations for general values of the parameters
to obtain the general boundary in parameter space that separates 
gap and no-gap solutions 
(section \ref{sec:5SDF}). We conclude in Section \ref{sec:7SDF}.

\section{Matrix current and Usadel equation}

\label{sec:1SDF}
Let us consider a $S/N/FI$ structure with 
the diffusive normal metal in the form
of a slab of length $L$ and cross-section $S$. 
This accounts both for sandwich $L \ll \sqrt{S}$ and wire
$L \gg \sqrt{S}$ geometries.
The Usadel equation  in the  normal part can be
presented as 
\begin{equation}
G_{N}\text{ }\frac{\partial }{\partial x}\left( \check{G}(x)\frac{\partial }{%
\partial x}\check{G}(x)\right) =-i\text{ }\frac{\tilde{G}_{Q}}{\delta }\left[
\varepsilon \text{ }\check{\tau}_{3},\check{G}(x)\right]  \label{Usadel}
\end{equation}
where $G_{N}=\sigma $ $S/L$ is the conductance associated with the diffusive
metal, $\sigma $ being its conductivity, $\tilde{G}_{Q}=e^{2}/\hbar $, $%
\delta \propto 1/SL$ is the average level spacing in the metal and $\varepsilon $ is the
energy of the quasiparticles (electrons and holes) with respect to the Fermi
energy $E_{F}$. $\check{G}(x)$ is an isotropic quasiclassical
Green's function in Keldysh$\otimes $Nambu$\otimes $Spin space, which is
denoted by $\left( \text{ }\stackrel{\vee }{}\mbox{
}\right) $. In Eq. (\ref{Usadel}), $x$ is the coordinate 
 normalized to the length L, $x=0(1)$ corresponding to the superconducting
(ferromagnetic) interface.

It is convenient to define the matrix current \cite{RefYuliFDS} as 
\begin{equation}
\frac{\check{I}}{G_{N}}=-\check{G}(x)\frac{\partial }{\partial x}\check{G}(x)
\label{Eq.diff}
\end{equation}
Substituting Eq. (\ref{Eq.diff}) into Eq. (\ref{Usadel}) we present 
the latter 
as conservation law of the matrix
current:  
\begin{equation}
\frac{\partial }{\partial x}\left( -\check{I}\right) =\check{I}%
_{leakage}(x)\rightarrow \check{I}_{0}-\check{I}_{1}=\int_0^1 dx\check{I}_{leakage}(x),
\label{Eqleakage}
\end{equation}
where ``leakage'' current $\check{I}_{leakage}(x)=-i$ $\tilde{G}_{Q}\left[ \varepsilon \text{ }%
\check{\tau}_{3}/\delta ,\check{G}(x)\right] $. 
 Note that the leakage current $\check{I}_{leakage}(x)$ does
not contribute to the nonequilibrium (physical) charge current, given
the fact that its Keldysh component is zero. 
This implies that the physical (charge)
current is conserved through the system, as expected. On the other hand, the
diagonal components of $\check{I}_{leakage}(x)$ in Keldysh space (Retarded
and Advanced components $\hat{I}_{leakage}^{R(A)}(x)$ ), which are
proportional to the energy $\varepsilon $, describe decoherence between
electrons and holes. 

By solving the Usadel equation in the normal metal, we obtain the
Green's function $\check{G}(x)$ that contains information about the
equilibrium spectral properties ($\hat{G}^{R(A)}(x)$) and about the
nonequilibrium transport properties ($\hat{G}^{K}(x)$). \cite{RammerFDS}
In this paper we concentrate on
spectral properties of the diffusive wire, so from now on we restrict
ourselves to the retarded block in Keldysh space. This is denoted by $\left( 
\text{ }\widehat{}\text{ }\right) $. Note that the retarded Green's function 
$\hat{G}^{R}(x)\equiv \hat{G}(x)$ is still a matrix of general structure
in Nambu$\otimes$Spin space.

If there is a single ferromagnetic element in the system, $\hat{%
G}(x)$ is diagonal in spin space and  can separated into two blocks for
spin parallel $(\uparrow )$ and antiparallel $(\downarrow )$ to the
magnetization of the ferromagnet. 
For each  spin component, $\hat{G%
}(x)$ can be parametrized in Nambu space as $\hat{G}(x)=\cos \theta (x)$ $%
\hat{\tau}_{3}+\sin \theta (x)$ $\cos \phi (x)$ $\hat{\tau}_{1}+\sin \theta
(x)$ $\sin \phi (x)$ $\hat{\tau}_{2}$, $\hat{\tau}_{1,2,3}$ being Pauli
matrices. If there is a  single superconducting reservoir
attached to the normal metal, 
the transport properties will not depend on
the absolute phase of the superconductor, so that 
$\phi (x)$ can be set to zero $%
\phi (x)=0$. Then $\hat{G}(x)$ depends on one parameter only: $\hat{G}(x)=\cos \theta (x)$ $\hat{%
\tau}_{3}+\sin \theta (x)$ $\hat{\tau}_{1}$. The phase (difference) $\phi (x)$ maybe important if the normal metal is connected to two or more superconducting reservoirs.

Using this parametrization for $\hat{G}(x)$, the retarded block of
Eqs.(\ref{Usadel}) and (\ref{Eq.diff}) transforms into a differential
equation for the angle $\theta (x)$: 
\begin{equation}
\frac{\partial ^{2}}{\partial x^{2}}\theta (x)+i\text{ }\frac{2\varepsilon }{E_{T}}\sin \theta (x)=0  \label{Usadelangle}
\end{equation}
and 
\begin{equation}
\hat{I}=-i\text{ }G_{N}\frac{\partial \theta (x)}{\partial x}\text{ }\hat{\tau}_{2}  \label{Iderivative}
\end{equation}
where we have introduced 
$E_{T}=\hbar D/L^{2}\equiv $ $G_{N}$ $\delta /\tilde{G}_{Q}$, the
Thouless energy associated with the normal metal.

\section{Boundary conditions}
\label{sec:2SDF}
\begin{figure}[t]
\begin{center}
\epsfig{file=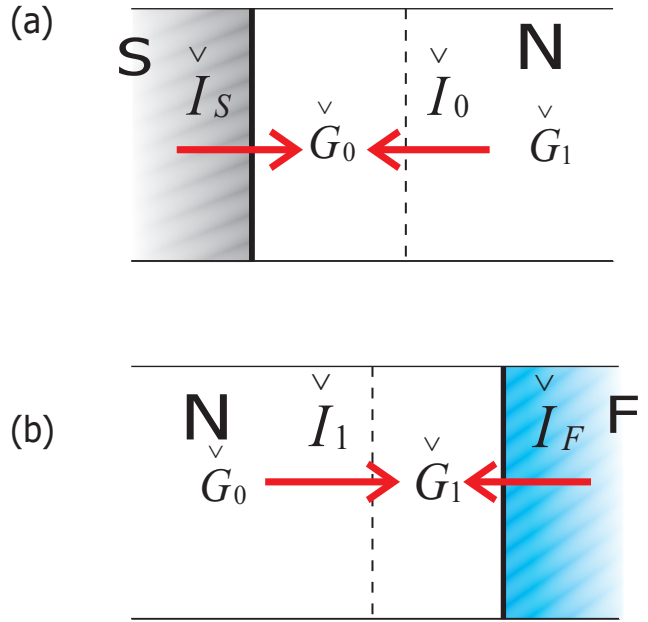,height=8.5 cm,width=8.5cm, clip=}
\end{center}

\caption{
Matrix current and boundary conditions.
Circuit-theory expressions give matrix currents
$\hat I_S, \hat I_F$ from the corresponding
reservoirs. These currents should match
matrix currents $\hat{I}_{0,1}$ from the Usadel equation,
defined via
derivatives of Green functions. This fixes
the boundary conditions for the Usadel equation.
}
\end{figure}

Circuit theory allows to find the boundary conditions at the 
interfaces  of
the normal metal in contact with the reservoirs 
simply by imposing matrix current conservation at these points. 
The matrix currents to the reservoirs are given by 
circuit theory expressions.
(FIG. 1).  We will assume that 
the superconducting reservoir is coupled to the normal metal 
through a tunnel contact. In addition, we disregard energy dependence
of Green functions in the reservoir assuming that the energy scale
of interest is much smaller than the superconducting energy gap $\Delta$
in the reservoir.  
Under these assumptions, the retarded Green function in the reservoir
is just $\hat \tau_1$.  The matrix current to the reservoir thus reads
\begin{equation}
\hat{I}_{S}=\frac{G_{S}}{2}\left[ \hat{\tau}_{1},\hat{G}(x)\right].
\end{equation}
To describe the matrix current to ferromagnetic insulator,
we use the results of our previous work \cite{DaniCondmat}
where we obtain 
\begin{equation}
\hat{I}_{F}=i\frac{G_{\phi }}{2}\left[ \vec{M}\text{ }\hat{\vec{\sigma}}%
\text{ }\hat{\tau}_{3},\hat{G}(x)\right] ,  \label{Iphi}
\end{equation}
$\vec{\sigma}$ being matrices in spin space, $\vec{M}$ being the magnetization
vector of the ferromagnet. 

The parameter $G_{\phi}$ has a dimension of conductance
and is related to the imaginary part of so-called mixing conductance
 $G_{\phi }=$Im$G^{\uparrow \downarrow }$. 
Mixing conductance has been introduced in Ref. \onlinecite{Braatas}
to describe the spin-flip of electrons reflected from
a  ferromagnetic boundary and is, in general, a complex
number. For insulating ferromagnet it is however
purely imaginary. In that case, $G_{\phi }$ acts as an effective magnetic field. Such  spin-dependent scattering situation is shown to be important
in magnetic insulator materials 
\cite{RefKumarPRL} and half-metallic magnets. 
\cite{CuevasSDF} 
Evaluation of $G_{\phi}$ for a simple model of
 insulating ferromagnet can be found
in Ref. \onlinecite{RefDaniPRLGBC}.

Using the parametrization in terms of $\theta (x)$, we find 
\begin{equation}
\hat{I}_{S}=-G_{S}\cos \theta_{0}\text{ }\hat{\tau}_{2}\text{}
\end{equation}
and 
\begin{equation}
\hat{I}_{FI}=\pm iG_{\phi }\sin \theta_{1} \text{ }\hat{\tau}_{2}\text{,}
\end{equation}
where the $+$($-$) sign corresponds to up (down) 
direction of spin with respect to the magnetization axis 
of the ferromagnet, $\theta_{0}=\theta (0)$ and $\theta_{1}=\theta (1)$.

Equating $\hat I_S = \hat I_0, \hat I_{FI} = \hat I_1$
gives the boundary conditions required,
\begin{equation}
-g_{S}\cos \theta _{0}=\frac{\partial \theta (x)}{\partial x}|_{0}
\label{IS}
\end{equation}
\begin{equation}
\pm ig_{\phi }\sin \theta _{1}=\frac{\partial \theta (x)}{\partial x}|_{1}
\label{IF}
\end{equation}
where we have intoduced two important dimensionless parameters characterizing
the stucture: 
$g_{S}=G_{S}/G_{N}$, $g_{\phi }=G_{\phi }/G_{N}$. 
As above, $\pm $ denotes two spin directions. 

The solutions of the Usadel equation  generally correspond
to complex  $\theta$. The density of states in a given point at
a given energy is $\nu(\epsilon,x) = \nu_0 \mbox{Re}[\cos(\theta(\epsilon,x))]$,
$\nu_0$ being the density of states in the absence of proximity effect.
In this paper, we concentrate on gap solutions where $\nu =0$ everywhere
in the normal metal. 

For this porpuse, it is convenient to define the complex angle 
$\theta (x)=\pi /2+i$ $\mu (x)$. Real $\mu(x)$ 
corresponds to gap solution. In terms of this angle,
the full system to solve reads
\begin{equation}
\frac{\partial ^{2}}{\partial x^{2}}\mu (x)+\tilde{\varepsilon}\cosh
\mu (x)=0  \label{Usadelmu}
\end{equation}
\begin{equation}
+g_{S}\sinh \mu _{0}=\frac{\partial \mu (x)}{\partial x}|_{0}
\label{BoundaryS}
\end{equation}
\begin{equation}
\pm g_{\phi }\cosh \mu _{1}=\frac{\partial \mu (x)}{\partial x}|_{1}
\label{BoundaryF}
\end{equation}
where we introduce the dimensionless energy 
$\tilde{\varepsilon}=2$ $\varepsilon /E_{T}$. 
The gap solutions exist in a certain region of
the three-dimensional parameter space $(g_S,g_{\phi},\tilde{\varepsilon})$.
To determine the boundary of this region is our primary task.

There are three limits where the solutions can be obtained 
analytically. The limit of vanishing resistance of the normal
metal, $ g_S \simeq g_{\phi} \simeq \tilde{\varepsilon}$,
can be treated with cirquit theory and has been considered in
Ref. \onlinecite{RefDaniPRLGBC}. Below we address the zero energy limit
($\tilde\varepsilon \approx 0$) and "spin-flip" limit 
($ g_S \simeq g^2_{\phi} \ll 1$).

\section{Zero energy}\label{sec:3SDF}
Here we will analyze the gap solutions at the Fermi level,
that is, at zero energy.
In this
case the leakage current conveniently dissapears, 
$\check{I}_{leakage}(x)=0$ and $\hat{I}$ = \emph{constant}. 
Eq. (\ref{Eq.diff}) can be easily integrated giving
$\hat G_1 = \exp(\hat I/G_{N}) \hat G_0$, where $\hat G_{0(1)}=\hat G(0(1))$. From this we obtain 
\begin{equation}
\hat{I}=-\frac{G_{N}}{2}\frac{\arccos \left( \text{\mbox{Tr}}\left( \hat{G}_{%
{\bf 0}}\text{ }\hat{G}_{{\bf 1}}\right) \right) }{\sqrt{4-\left( \text{%
\mbox{Tr}}\left( \hat{G}_{{\bf 0}}\text{ }\hat{G}_{{\bf 1}}\right) \right)
^{2}}}\left[ \hat{G}_{{\bf 0}},\hat{G}_{{\bf 1}}\right],
\end{equation}
which can be further simplified by 
using the parametrization $\hat{G}_{{\bf 0(1)}}=\cos \theta _{{\bf 0(1)}}$ $\hat{\tau}%
_{3}+\sin \theta _{{\bf 0(1)}}$ $\hat{\tau}_{1}$ 
into the following simple expression 
for the current through the diffusive normal metal:
\begin{equation}
\hat{I} =-i\ G_{N}(\theta _{0}-\theta _{1})\hat{\tau}_{2} = 
\ G_{N}(\mu_{0}-\mu _{1})\hat{\tau}_{2}. 
\label{currentangle}
\end{equation}
Taking into account the boundary conditions on both interfaces,
we obtain the equations for $\mu_{0,1}$,
\begin{equation}
g_{S}\sinh \mu _{0}=\mu _{1}-\mu _{0} =\pm g_{\phi }\cosh \mu _{1}.  
\label{gsmu}
\end{equation}

\begin{figure}[t]
\begin{center}
\epsfig{file=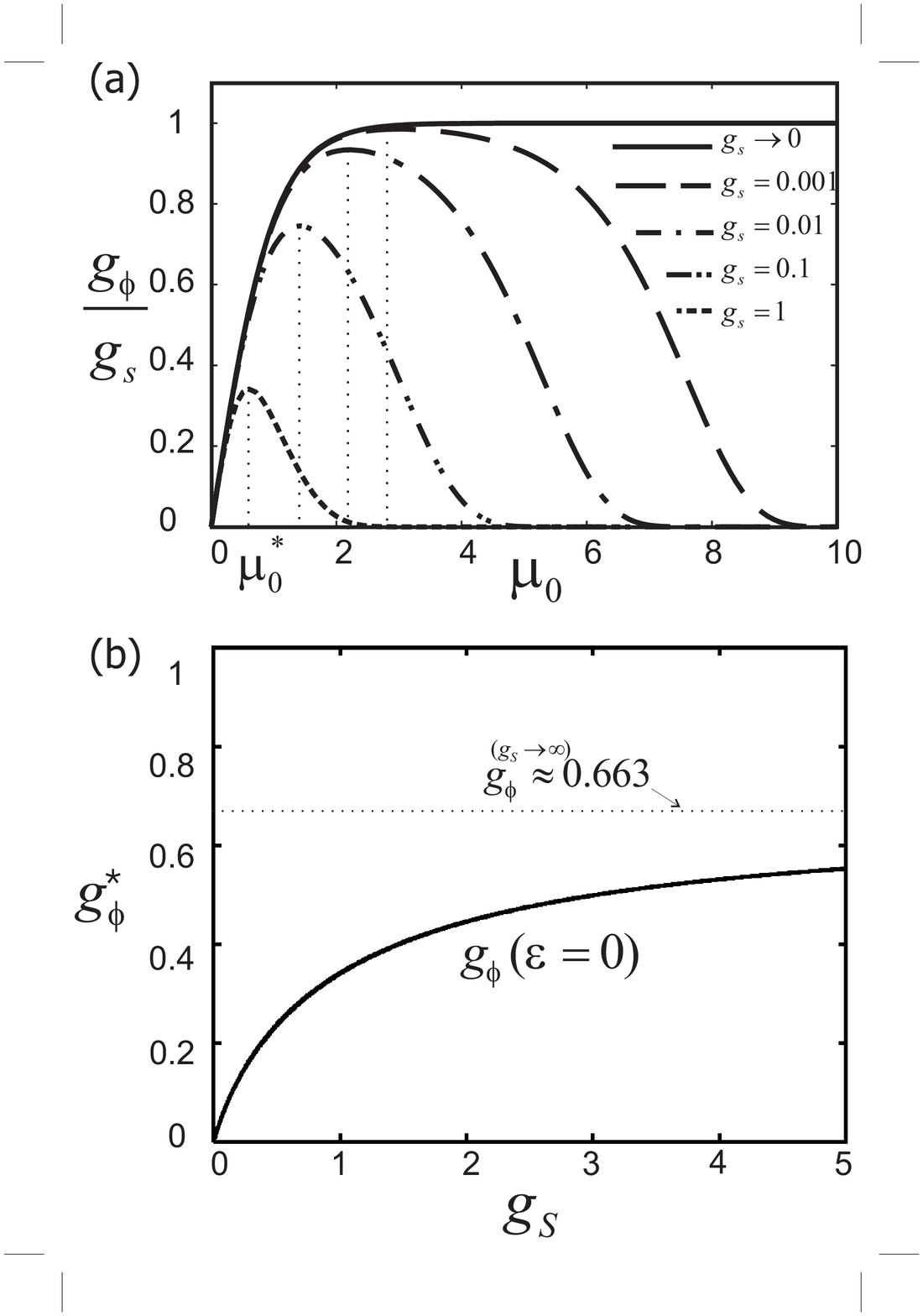,height=10.5 cm,width=8.5cm ,clip=}
\end {center}
\caption[]{\label{fig:Fig2SDF}(a) $g_{\phi }/g_{S}$ versus
$\mu _{0}$ for various values of $g_{S}$ (Eq.\ref{gsgphi}). 
The maximum achieved at  $\mu _{0}^{\ast }$
gives the maximum $g_{\phi}$ at which the gap persists at given 
$g_{S}$.
(b) The maximum $g_{\phi}^{\ast }$ versus $g_{S}$.
The curve saturates at $g_{\phi} \approx 0.663$ for $g_{S} \to \infty$.}
\end{figure}

We can readily express from these two equations $g_{\phi
}/g_{S}$ as a function of $\mu _{0}$ and $g_{S}$ 
\begin{equation}
\label{gsgphi}
\frac{g_{\phi }}{g_{S}}=\frac{\sinh \mu _{0}}{\cosh \left( \mu
_{0}+g_{S}\sinh \mu _{0}\right) }=f(\mu _{0},\text{ }g_{S})  
\end{equation}
Here we concentrate on the spin-up component. 
The solution for spin-down component corresponds to different sign of $\mu_0$.
In FIG. 2a, we plot $g_{\phi }/g_{S}$ versus
$\mu _{0}$ for several values of $g_{S}$. 
We see that $g_{\phi }/g_{S}$ reaches a maximum value $\left( g_{\phi
}/g_{S}\right) ^{\ast }$ at a certain value $\mu _{0}^{\ast }$.
 The position
and the height of the maximum changes by changing $g_{S}$,
$\mu(0)^{\ast} \to \infty, (g_{\phi}/g_S)^{\ast} \to 1$ if $g_S \to 0$. 

So for a given $g_{S}$, the maximum possible value of $g_{\phi }$ such as
there is still a gap in the induced density of states of the wire is given
by 
\begin{equation}
g_{\phi }^{\ast }=f(\mu _{0}^{\ast },g_{S})g_{S}.  \label{gmaximum}
\end{equation}
In FIG. 2b we plot $g_{\phi }^{\ast }$ as a function of $g_{S}$. 
This curve defines the boundary between gap 
(below) region and no gap
(above) region in $g_{\phi }-g_{S}$ 
parameter space at zero energy $\varepsilon =0$.
As expected, magnetic correlations combat the proximity effect
at the Fermi level and the gap solutions dissapear upon increasing
$g_{\phi}$. Even if the coupling to the superconductor is infinitely
strong, $g_S \to \infty$, the gap survives only if $g_{\phi} < 0.663$.

Let us expand  
$g_{\phi }/g_{S}$ near its maximum value $\left( g_{\phi }/g_{S}\right)
^{\ast }$. Defining deviations from this point $\Delta g_{\phi -S}=\left( g_{\phi }/g_{S}\right)
-\left( g_{\phi }/g_{S}\right) ^{\ast }$ and $\bar{\mu}=\mu _{0}-\mu
_{0}^{\ast }$, the expansion is obviously
\begin{equation}
\Delta g_{\phi -S}=-\text{ }C\text{ }\bar{\mu}^{2}  \label{gphigsmax1}
\end{equation}
 $C$ being a positive constant. Let us note that the density
of states 
$\nu(0)/\nu_0 = \mbox{Im} \ \sinh(\mu_0) \propto \mbox{Im} \bar \mu$.
This gives a square-root singularity of the density of states near the 
boundary,
\begin{equation}
\nu \propto \sqrt{\Delta g_{\phi -S}}. \label{nusqrt}
\end{equation}
The limit $g_{S}\rightarrow 0$ deserves some special consideration.
It corresponds to the circuit-theory limit where 
the diffusive normal metal  
simply reduces to a 
``node" with spatially independent 
Green's function $\check{G}$. 
This gives however, a BCS-like 
{\em inverse} square-root singularity
in the density of states at the boundary given at $g_S=g_{\phi}$, 
$\nu/\nu_0 = 1/\sqrt{1 -(g_{S}/g_{\phi})^2 } \approx 1/
\sqrt{2 \Delta g_{\phi -S}}$ at $\Delta g_{\phi -S}\ll 1$.
This result looks difficult to reconcile with the result obtained in Eq.(\ref{nusqrt}).

The point is that for $g_S \ll 1$ 
 an extra crossover takes place such the density of states changes from inverse square root to square root. The crossover occurs close to the gap boundary $g_S=g_{\phi}$. From Eq. (\ref{gsgphi}) it is clear that for $g_S \ll 1$, the values of $g_{\phi}/g_S \simeq 1$  occur at $\mu_0 \gg 1$. Close to the boundary, an expansion of Eq. (\ref{gsgphi}) up to terms $\sim \exp({-2\mu_0})$ is possible. The evaluation of the maximum of $g_{\phi}/g_S$ as function of $\mu_0$ allows to determine the constant $C$ in Eq. (\ref{gphigsmax1}), $C \simeq g_S^{4/3}$. The crossover occurs then at $\Delta g_{\phi -S} \simeq  g_S^{2/3} \ll 1$.
Below the crossover, at 
$\Delta g_{\phi -S} \ll g_S^{2/3}$,
there is a square-root singularity 
$\nu/\nu_0 \propto g^{-2/3}_{S} \sqrt{\Delta g_{\phi -S}}$
that changes to
$\nu/\nu_0 \simeq 1/\sqrt{2 \Delta g_{\phi -S}}$
above the crossover, 
at 
$\Delta g_{\phi -S} \gg g_S^{2/3}$.
The maximum density of states is 
therefore $\nu/\nu_0 \simeq g_S^{-1/3}$. 

\section{"Spin-flip" limit}
\label{sec:4SDF}

Now we would like to extend the results of the previous section
to finite energy $\varepsilon$.
We do this assuming that the Green function
does not change much along the normal metal, 
so that $|\theta_1 -\theta_0|
\ll \theta_0$.

We start again with 
Eqs. (\ref{Usadelangle}), (\ref{IS}) and (\ref{IF}) . 
Integrating  Eq.(\ref{Usadelangle}) over $x$ and  
using Eqs.(\ref{IS}) and (\ref{IF}) gives 
\begin{equation} 
\pm ig_{\phi }\sin \theta _{1}+g_{S}\cos \theta _{0}+i\text{ }\tilde{%
\varepsilon}\int_{0}^{1}dx\sin \theta (x)=0.  \label{Eqintbound}
\end{equation}
If we assume that the Green function does not change with $x$,
$\theta(x) = \theta_0$, we derive from Eq. ({\ref{Eqintbound}})
the circuit-theory equation
\begin{equation}
i\left( \tilde{\varepsilon}\text{ }\pm g_{\phi }\right) \sin \theta
_{0}+g_{S}\cos \theta _{0}=0. \label{AG0}
\end{equation}
With this, 
we reproduce the results dicussed in Ref.\onlinecite{RefDaniPRLGBC} 
: the density of states mimics 
the one of a BCS superconductor in the presence of 
the spin-splitting magnetic field   
\begin{equation}
\upsilon (\tilde{\varepsilon} )=\frac{\left| \tilde{\varepsilon}
\pm g_{\phi}\right| }{\sqrt{(\tilde{\varepsilon} \pm g_{\phi})^{2}-g_{S}^{2}}}.  \label{DOS}
\end{equation}
The presence of a gap is strictly speaking a non-perturbative effect. The perturbation expansion of the density of states shown in Eq. (\ref{DOS}) is valid at high energies $|\varepsilon \pm \tilde{h}| \gg \tilde{\Delta}$. The  leading ``spin-dependent'' correction is proportional to $(\tilde \Delta^{2} \tilde{h})/\varepsilon^{3} \equiv ( g_{S}^{2} g_{\phi})/\tilde{\varepsilon}^{3}$. This implies that the leading diagram in perturbation series involves four tunneling amplitudes at the $S/N$ interface and two spin-dependent reflection amplitudes at the $N/FI$ interface.  

The equation (\ref{AG0}) is correct in the leading order in
$\tilde \varepsilon, g_S, g_{\phi} \ll 1$.
There can be however a problem if $\tilde \varepsilon$ is
too close to $\mp g_{\phi}$ since the first coefficient
is anomalously small in this case. To account for this,
we should re-derive these equations to the next-to-leading order. 

We present $\theta(x)$ in the form that satisfies boundary
conditions,
\begin{equation}
\theta (x)=\theta _{0}+\left( \theta _{1}-\theta _{0}\right) x+\theta
^{(1)}(x);\;  \theta^{(1)}(0)= \theta^{(1)}(1)=0\label{thetaanzats}
\end{equation}
and evalutate the corrections $\theta_{1}-\theta_{0}, \theta^{(1)}(x)$
in the leading order. To evaluate $\theta_{1}-\theta_{0}$,
let us multiply Eq.(\ref{Usadelangle}) by $x$ 
and integrate by parts the first term $\int_{0}^{1}dx$ $x$ $\ddot{\theta}(x)$ where ($\ddot{%
\theta}(x)\equiv \partial ^{2}\theta (x)/\partial x^{2}$), obtaining the
following expression 
\begin{equation}
\dot{\theta}_{1}-\left( \theta _{1}-\theta _{0}\right) +i\text{ }\tilde{%
\varepsilon}\int_{0}^{1}dx\text{ }x\text{ }\sin \theta (x)=0.  \label{trick1}
\end{equation}
We set $\theta(x) = \theta_0$ under the sign of integral
and use  Eq. (\ref{IF}) to obtain  
 \begin{equation}
\theta _{1}-\theta _{0}=i\left( \pm g_{\phi }\text{ }+\frac{\tilde{%
\varepsilon}}{2}\right) \sin \theta _{0}.  \label{trick3}
\end{equation}
With the same accuracy, the differential equation for 
$\theta ^{(1)}(x)$ reads
\begin{equation}
\frac{\partial ^{2}}{\partial x^{2}}\theta ^{(1)}(x)+i\text{ }\tilde{%
\varepsilon}\sin \left( \theta _{0}\right) =0.  \label{usadeltilde1}
\end{equation}
This results in 
\begin{equation}
\theta ^{(1)}(x)\approx i\text{ }\tilde{\varepsilon}\frac{\sin \theta _{0}}{2%
}(x-x^{2}). \label{theta1x}
\end{equation}
Finally we substitute Eqs. (\ref{theta1x}) to Eq. (\ref{Eqintbound}) to get
the following relation:
\begin{equation}
i\left( \tilde{\varepsilon}\text{ }\pm g_{\phi }\right) \sin \theta
_{0}+g_{S}\cos \theta _{0}- \zeta g_S \sin \theta _{0}\cos
\theta _{0}=0,  \label{AG1}
\end{equation}
where $\zeta =\left( g_{\phi }^{2}\pm g_{\phi }\tilde{\varepsilon}+\tilde{%
\varepsilon}^{2}/3\right)/g_S$. As we mentioned above, $\zeta$ plays a role
only if $|\tilde{\varepsilon}\text{ }\pm g_{\phi }| \ll \tilde{\varepsilon},g_{\phi}$.
Under these conditions, the energy dependence of $\zeta$ can be disregarded and
$\zeta = g_{\phi }^{2}/3g_S$.

The relation (\ref{AG1}) resembles very much one of the most
important equations in the superconductivity theory that was
first derived by Abrikosov and Gor'kov \cite{AG} 
to describe suppresion of superconductivity by magnetic impurities.
Precise association is achieved by the following change of notations:
\begin{eqnarray*}
(\tilde \varepsilon \pm g_{\phi})E_{T} &\to& E,\\
g_S E_{T} &\to &\Delta,\\
\zeta\Delta =  E_T g^2_{\phi}/3 &\to & 1/\tau_{s}  
\end{eqnarray*}
where $E,\Delta,\tau_{s}$ are respectively 
energy, superconducting order parameter and spin-flip time due to
magnetic impurities.\cite{AG} This is why we refer to the limit under consideration
as to "spin-flip" limit. To remind, there are no magnetic impurities 
in the structure considered, and effective spin-flip comes from
interplay of diffusion in the normal metal and reflection at
the $N/FI$ interface. 
 
Maki has demonstrated that similar equation
accounts for gapless superconductivity in a variety of
circumstances, $\zeta$ being the pair-breaking parameter.\cite{MakiFDS} 
We define $\theta _{0}=\pi /2+i$ $\mu _{0}$, $u=\tanh \mu _{0}$, 
$\omega =\left( \tilde{\varepsilon}\text{ }\pm g_{\phi }\right)/g_{S} $,
to be close to notations of Ref. \onlinecite{MakiFDS}. This gives
\begin{equation}
\omega =u\left( 1-\zeta \frac{1}{\sqrt{1-u^{2}}}\right)
\label{MAKIECU}
\end{equation}
The maximum value of $\omega$ with respect to real $u$, $\omega^{\ast}$
gives the
energy interval around $\tilde\varepsilon = \mp g_{\phi}$ 
where the gap solutions occur.
This value is determined by the condition 
$\partial \omega /\partial u=0$, which gives
\begin{equation} \omega ^{\ast
}=\text{ }\left( 1-\zeta ^{2/3}\right) ^{3/2} \label{w*}
\end{equation}
achieved at $u=u^{*}$,
 \begin{equation} u^{\ast }=\left( 1-\zeta ^{2/3}\right)
^{1/2}. \label{u*} \end{equation}
 
\begin{figure}[ht] \begin{center} \epsfig{file=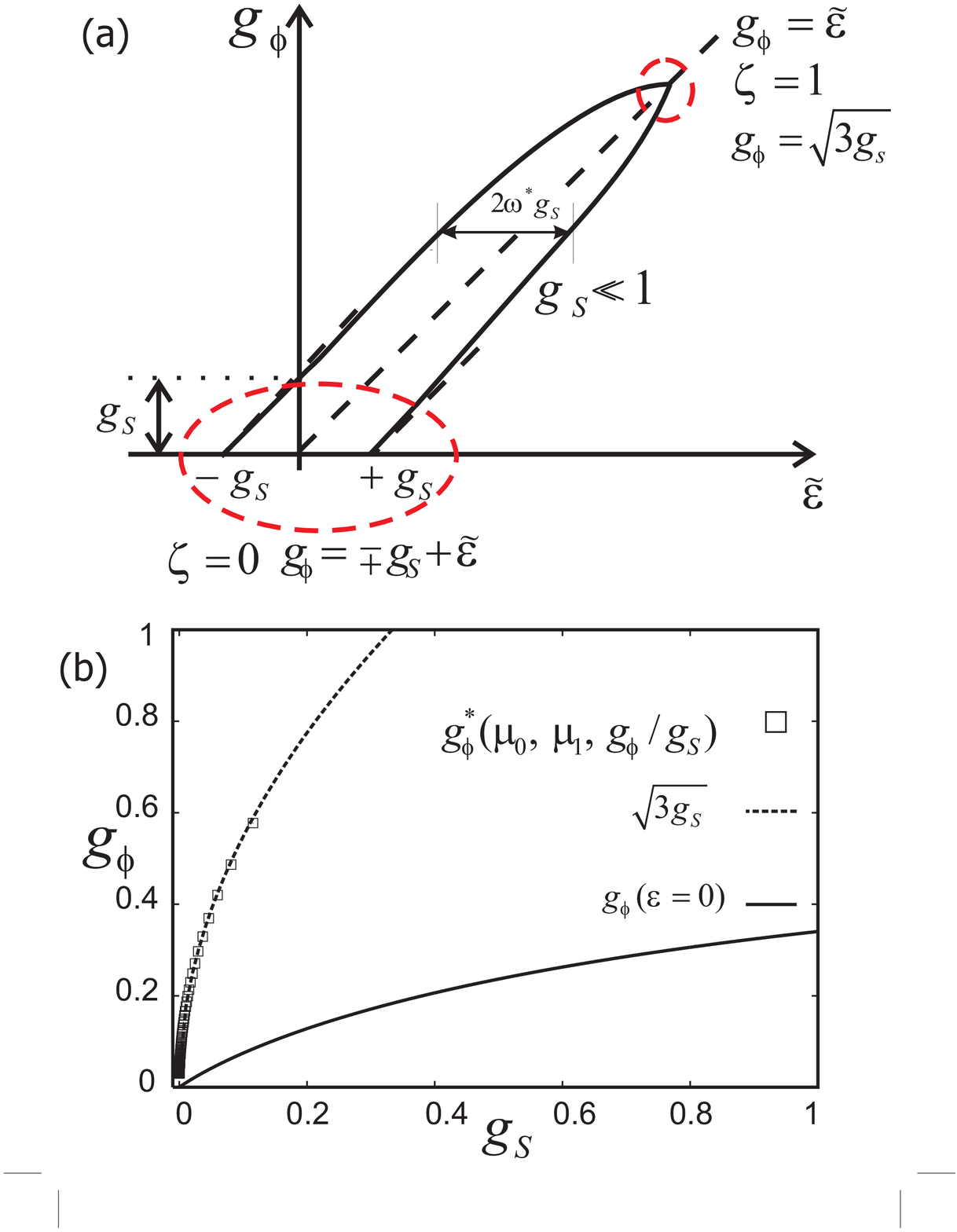,height=10.5
cm,width=8.5cm ,clip=} \end {center} 
\caption[]{\label{fig:Fig3SDF} (a)
The sketch of the gap domain in $\tilde\varepsilon - g_{\phi}$ plane
at $g_{S} \ll 1$.
 (b) Exact results (squares) for the maximum $g_{\phi}$ at $\zeta=1$, 
follow the expected $\sqrt{3 g_{S}}$ dependence for small $g_{S}$. The lower curve is the one plotted in FIG. 2. } 
 \end{figure} 

There are no gap solutions if $\zeta > 1$.
The region where these solutions do occur is sketched in FIG. 3a in
$\tilde\varepsilon-g_{\phi}$ coordinates. It looks like a $45^{\circ}
$ slanted strip, the width of the stip in horizontal direction being
given by $2 \omega^{\ast} g_{S}$. Near the origin $\zeta=0$. In this situation, it is obtained from 
Eqs.(\ref{w*}, \ref{u*}) that the width is $2g_{S}$ and $|\tilde\varepsilon \pm g_{\phi}|= g_{S}$. 
The gap solutions at zero energy $\tilde\varepsilon=0$ dissapear at $g_{\phi} = g_{S}$. 
The width gradualy reduces with increasing 
$g_{\phi}$ due to the increase of $\zeta$. The strip ends if $\zeta=1$, that is, at
$|g_{\phi}| = \sqrt{3 g_S} \gg g_{S}$ (FIG. 3b). 

This demonstrates that even in the limit of
$g_{\phi},g_{S} \ll 1$ the ferromagentic insulator 
not only shifts the gap states but also reduces 
and finally suppesses the gap due to effective spin-flip.
We show in the next section that the same picture 
is qualitatively valid for arbitrary values of these parameters.
 
\section{gap-no gap boundary in general case}

\label{sec:5SDF}

So far we have studied the boundary separating gap and no-gap
solutions in the parameter space for two limiting
cases that allow for analytic solutions. 
In this section, we find this boundary for
arbitrary values of the parameters. We do this by
solving Eqs. (\ref{Usadelmu},\ref{BoundaryS},\ref{BoundaryF}) 
numerically.

\begin{figure}[t]
\begin{center}
\epsfig{file=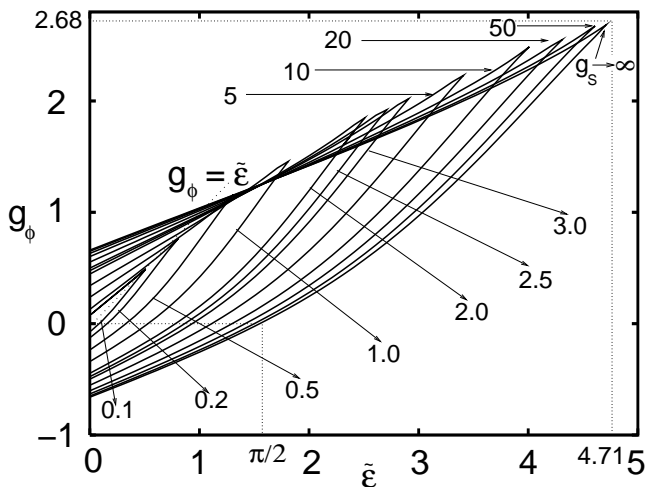,height=6.5 cm,width=8.5cm} 
\end {center}
 
\caption[]{
The gap domains (strips) in $g_{\phi }-\tilde{\varepsilon}$ plane
calculated for different values of $g_{S}=[0.1, 0.2,.., 50, \infty]$. Their shape is similar
to the sketch in FIG. 3a. 
The tip of each
strip gives the maximum value of $g_{\phi}$ at which the gap survives
and simultaneously the 
corresponding energy.
}

\end{figure}

\begin{figure}[t]
\begin{center}
\epsfig{file=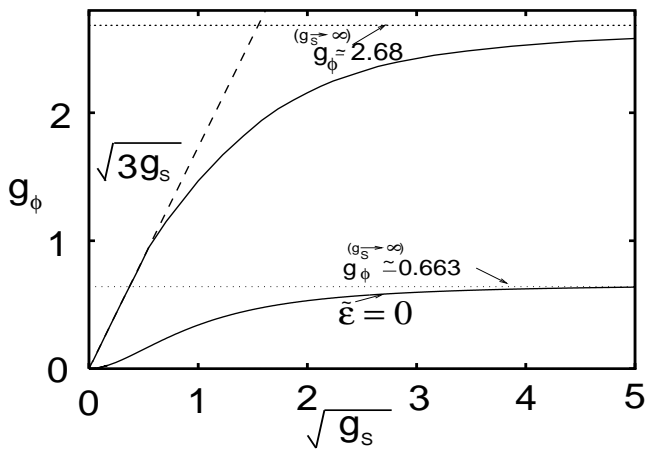,height=6.0 cm,width=8.5cm ,clip=}
 
\end {center}
 
\caption[]{
Side view of the boundary surface in three-dimensional ($g_{S},g_{\phi},\tilde\varepsilon$) space. The upper curve gives the maximum
$g_{\phi}$ connecting the tips of the strips plotted in FIG.4.
It rises monotonically with $g_{S}$
to reach the asymptotic value $\simeq 2.68$ at
$g_{S}\rightarrow \infty$.
The lower  curve presents a cross-section of the surface
with the $\tilde\varepsilon=0$ plane.} 
\end{figure}

Since we only concern with the boundary, 
the numerical procedure is as follows.
We fix $g_{S}$ and $\tilde \varepsilon$. The solutions 
of Eq. \ref{Usadelmu} with boundary condition
(\ref{BoundaryS}) can be parametrized with $\mu_0$.
We express $\mu_1,\dot \mu_1$ in terms of $\mu_0$.
Then the last boundary condition (\ref{BoundaryF})
could be solved to find $\mu_0$ in terms of $g_{\phi}$.
We do the opposite: We use (\ref{BoundaryF}) to directly express 
$g_{\phi}$ in terms of $\mu_0$
and find the two extrema of $g_{\phi}(\mu_0)$.
Those  give the endpoints of the interval
of $g_{\phi}$ where the gap solutions exist --- elements of the
boundary. 
We plot these endpoins at fixed $g_{S}$ versus dimensionless
energy $\tilde \varepsilon$ to obtain slanted strips similar
to the one in FIG. 3a. At certain energy, the extrema come together
indicating the endpoint of the strip. 
 
In FIG. 4 we show these strips
in $g_{\phi} - \tilde \varepsilon$ plane
for a wide range of values of $g_{S}$. As expected from the previous
discussion, for $g_{S}\ll 1$ the strips  
extend along the
$g_{\phi}=\tilde{\varepsilon}$ line. 
The sharp tip of each strip
gives the critical 
value of $g_{\phi}$ at which for a given $g_{S}$ 
the induced minigap  disappears. For small $g_S$, the height
of the tip, $\sqrt{3 g_S}$, is much bigger than the width
of the strip $g_S$.
With increasing $g_{S}$,
the shape of the strips changes. 
They increase both in width and height, so that these dimensions
become of the same order. The strips also become less slanted.
The shape converges at $g_{S} \to \infty$ (outer curve in FIG.4).
In this limit, the maximum  
$g_{\phi }$ that allows for superconductivity 
is $ \approx 2.68$ and is achieved at 
 $\tilde{\varepsilon} \approx 4.71$. 
It is interesting to note that this energy is 
higher than the maximum value of the minigap without
magnetic correlations
($\tilde\varepsilon(g_{\phi}=0) = \pi/2,$ see FIG. 4).
Counterintuitevely, the presence of the magnetic insulator
helps the gap solutions to persist at higher energy.
Albeit the magnetic correlations quickly remove these solutions 
from the Fermi level.

Each strip is a cross-section of the boundary surface
in three-dimensional
$(g_{\phi},g_{S},\tilde\varepsilon)$ space.
We present in FIG. 5 the side view of this surface.
The lower curve in this figure is the cross-section
of the surface with $\tilde\varepsilon=0$ plane
and shows the critical value of $g_{\phi}$ at which
the gap solutions dissapear from the Fermi level.
The same curve has been already presented in FIG. 2.
The upper curve gives the critical value of $g_{\phi}$
at which gap solutions dissapear at any energy.
It consists of the tips of each strip from FIG. 4 as 
a function of $g_{S}$. 
We see that at $g_{S} \to \infty$ this curve satures
at $g_{\phi}= 2.68$. The asymptotics $g_{\phi} =\sqrt{3
g_{S}}$ derived in the previous section agree with this curve 
at $g_{\phi} < 1$ as expected.

\section{Conclusions}

\label{sec:7SDF}
We have studied the proximity effect in $S/N/FI$ structures
with N being a diffusive normal metal. We pay special attention
to the gap in the density of states
and find its domain in the
parameter space. The convenient dimensionless parameters
to work with are $g_{\phi},g_{S}$ that characterize the
intensity of magnetic and superconducting correlations
respectively, and energy in units of Thouless energy,
$\tilde \varepsilon$.

We demonstrate that the combined effect of a ferromagnetic
insulator and the elastic scattering on the proximity gap of a diffusive wire is twofold. 
First, the ferromagnetic insulator  provides an effective exchange field $\tilde{h}$ that
shifts the gap edges in opposite directions for
opposite components without reducing the energy
interval where the gap solutions occur.
Second, its effect combined with suficiently strong elastic scattering in N  
reduces this interval and 
finally suppresses the gap. In the limit
of small $g_{S},g_{\phi}$
("spin-flip" limit) the mechanism
of this suppression is precisely equivalent to the known one
from magnetic impurities, with an effective spin-flip
rate $1/\tau_{s} = g_{\phi}^2 E_{T}/3$. 
Qualitatively, this picture remains valid
at arbitrary parameters.

If $g_{\phi} > 0.66$ no gap persist at the Fermi level.
If $g_{\phi} > 2.68$ no gap occurs at any energy.
Counterintuitively, the gap in the presence of
magnetic correlations may occur at energies higher
than in the absence of the ferromagnetic insulator.

The absence or the presence of the gap in the normal 
part of a $S/N/FI$ structure at certain energy 
can be observed by a spin-sensitive tunnel probe.
The resistance of such probe must exceed all interface resistances.
The possible implementation of the probe depends on its concrete
geometry. For traditional sandwich geometry, it is probably
simpler to keep the $FI$ layer rather thin, so that the electrons
can tunnel through. Covering this layer with a conducting 
ferromagnet makes the tunnel probe. For the wire geometry,
small tunnel contacts to ferromagnets can be made in different
points of the normal wire. An alternative is the suggestion
of Ref. \onlinecite{RefDaniPRLGBC}: the tunneling between
two $FI/N/S$ structures.

We thank W. Belzig, D. Esteve and Ya. M. Blanter for useful discussions. 
This work was financially supported by the Stichting voor 
Fundamenteel Onderzoek der Materie (FOM). 
D. H-H also acknowledges additional financial support from the  U.S. DOE grant DE-FG-0291-ER-40608.



\begin{thebibliography}{99}
\bibitem{DeGennes}
G. Deutscher and P.G. de Gennes, in {\it Superconductivity}, edited by R.D.
Parks (Dekker, New York, 1969), p. 1005;P.G. de Gennes, Phys. Lett. {\bf 23}, 10 (1969);
G. Deutscher and F. Meunier, Phys. Rev. Lett. \textbf{22},395 (1969);
J. J. Hauser Phys. Rev. Lett. \textbf{23},374 (1969);
\bibitem{pi}
V.V. Ryazanov, V.A. Oboznov, A.Y. Rusanov, A.V. Veretennikov,
A.A. Golubov, and J. Aarts, Phys. Rev. Lett. {\bf 86}, 2427 (2001);
T. Kontos, M. Aprili, J. Lesueur, F. Genet, B. Stephanidis, and 
R.Boursier, Phys. Rev. Lett. {\bf 89}, 137007 (2002).
\bibitem{spinvalve} A.I. Buzdin, A.V. Vedyayev, and N.V. Ryzhanova,
Europhys. Lett. {\bf 48}, 686 (1999);
L. R. Tagirov,  Phys. Rev. Lett. 83, 2058 (1999).
\bibitem{triplet} A. F. Volkov, F. S. Bergeret and K. B. Efetov, 
Phys. Rev. Lett. {\bf 90}, 117006 (2003);
F. S. Bergeret, A. F. Volkov, and K. B. Efetov, 
Phys. Rev. Lett. {\bf 86}, 4096 (2001).
\bibitem{trilayers} R. M\'{e}lin and S. Peysson, Phys. Rev. B {\bf 68}, 174515 (2003); 
R. M\'{e}lin, Eur. Phys. J B {\bf 39}, 249 (2004); R. M\'{e}lin and D. Feinberg, cond-mat/0407283.
\bibitem{entanglement} G. Falci, D. Feinberg, and F. W. J. Hekking, Europhys.
Lett. 54, 225 (2001); N. M. Chtchelkatchev, JETP Lett. 78, 230 (2003).

\bibitem{RefTedrowPRL} P.M. Tedrow and R. Meservey, Phys. Rev. Lett. 
{\bf 27}, 919 (1971); 
R. Meservey and P.M. Tedrow, Phys. Rep. {\bf 238}, 173 (1994); 
 J. Y. Gu, C.-Y. You, J. S. Jiang, J. Pearson, Ya. B. Bazaliy, and 
S. D. Bader, Phys. Rev. Lett. \textbf{89},267001 (2002). 


\bibitem{Gapless} R. Fazio and C. Lucheroni, Europhys.
Lett. {\bf 45}, 707 (1999); K.  Halterman and O. T. Valls, Phys. Rev.
B {\bf 65}, 014509 (2001); ibid. B {\bf 66}, 224516 (2002); B {\bf 69}, 014517
(2004); F. S. Bergeret, A. F. Volkov and K. B. Efetov, Phys. Rev. B {\bf 65}, 134505 (2002).

\bibitem{AG} A. A. Abrikosov, L. P. Gorkov
Sov. Phys. JETP-USSR {\bf 12}, 1243 (1961).

\bibitem{WeertArnold} 
M.J. DeWeert and G.B. Arnold, Phys. Rev. Lett. {\bf 55}, 1522 (1985); 
M.J. DeWeert and G.B. Arnold, Phys. Rev. B {\bf 39}, 11307 (1989). 

\bibitem{RefTokuyasu} T. Tokuyasu, J.A. Sauls and D. Rainer, Phys. Rev. B {\bf 38}, 8823 (1988).

\bibitem{RefKumarPRL}
P. M. Tedrow, J. E. Tkaczyk and A. Kumar, Phys. Rev.
Lett. {\bf 56}, 1746 (1986).


\bibitem{RefDaniPRLGBC}
D. Huertas-Hernando, Yu. V. Nazarov and W. Belzig,
Phys. Rev. Lett. \textbf{88},047003 (2002).

\bibitem{AndreevINTRO}  A. F. Andreev, Sov. Phys. JETP \textbf{19},1228
(1964).

\bibitem{TinkhamINTRO}  M. Tinkham, \emph{Introduction to Superconductivity}%
, 2nd ed., (McGraw Hill, N.Y., 1996).

\bibitem{KulikINTRO}  I. O. Kulik, Sov. Phys. JETP \textbf{30}, 944 (1970).

\bibitem{RefMcMillanPRL}  W.L. McMillan, Phys. Rev. \textbf{175}, 537 (1968).

\bibitem{RefMcMillan1PRL} S. Gu\'{e}ron, H. Pothier, Norman O. Birge, D. Esteve, and M. H. Devoret , Phys. Rev. Lett. \textbf{77}, 3025(1996); W. Belzig, C. Bruder and G. Sch\"{o}n, Phys. Rev. B \textbf{54}, 9443 (1996); E. Scheer, W. Belzig, Y. Naveh, M. H. Devoret, D. Esteve, and C. Urbina , Phys. Rev. Lett. \textbf{86}, 284 (2001); N. Moussy, H. Courtois and B. Pannetier, Europhys. Lett.\textbf{55}, 861 (2001).

\bibitem{PaulSDF}  A. A. Golubov and M. Yu. Kupriyanov, Sov Phys. JETP {\bf %
69}, 805 (1989); A. Lodder and Yu. V. Nazarov, Phys. Rev. B {\bf 58}, 5783 (1998); S. Pilgram, W. Belzig and C. Bruder, Phys. Rev. B {\bf 62}, 12462 (2000); P. M. Ostrovsky, M. A. Skvortsov and M. V. Feigel'man, Phys. Rev. Lett. {\bf 87}, 027002 (2001).
\bibitem{UsadelFDS}
K. D. Usadel, Phys. Rev. Lett. {\bf 25}, 507 (1970).

\bibitem{RefZaitsev} A.V. Zaitsev, Sov. Phys. JETP {\bf 59}, 1015 (1984).




\bibitem{RefSauls} A. Millis, D. Rainer and J.A. Sauls, Phys. Rev. B {\bf 38}, 4504 (1988). 

\bibitem{RefYuliFDS}  Yu. V. Nazarov, Phys. Rev. Lett. {\bf 73}, 1420 (1994);
Yu.V. Nazarov, Superlatt. and Microstruc. {\bf 25}, 1221 (1999).

\bibitem{RefBCKopu} J. Kopu, M. Eschrig, J.C. Cuevas, and M. Fogelstrom, Phys. Rev. B {\bf 69}, 094501 (2004).

\bibitem{Zareyan} M. Zareyan, W. Belzig and Yu. V. Nazarov, Phys. Rev. Lett. \textbf{86}, 308 (2001).


\bibitem{RammerFDS}  J. Rammer and H. Smith, Rev. Mod. Phys. {\bf 58}, 323
(1986).
\bibitem{DaniCondmat} D. Huertas-Hernando, W. Belzig and Yu. V. Nazarov, cond-mat/0204116.
\bibitem{Braatas} A. Brataas, Yu. V. Nazarov and G.E.W. Bauer, Phys. Rev.
Lett. {\bf 84}, 2481 (2000).

\bibitem{CuevasSDF}  M. Eschrig, J. Kopu, J. C. Cuevas and G. Sch\"{o}n, Phys. Rev. Lett. {\bf 90}, 137003 (2003). 

\bibitem{MakiFDS}  K. Maki {\em Gapless Superconductivity}, chap. 18, Vol. 2
in {\em Superconductivity}, edited by R.D. Parks (Marcel Dekker, New York,
1969).
\end{thebibliography}
\end{document}